# Marvista: Exploring the Design of a Human-AI Collaborative News Reading Tool

XIANG 'ANTHONY' CHEN, UCLA HCI Research,
CHIEN-SHENG WU, Salesforce Research,
LIDIYA MURAKHOVS'KA, Salesforce Research,
PHILIPPE LABAN, Salesforce Research,
TONG NIU, Salesforce Research,
WENHAO LIU, Faire,
CAIMING XIONG, Salesforce Research,

We explore the design of Marvista—a human-AI collaborative tool that employs a suite of natural language processing models to provide end-to-end support for reading online news articles. Before reading an article, Marvista helps a user plan what to read by filtering text based on how much time one can spend and what questions one is interested to find out from the article. During reading, Marvista helps the user reflect on their understanding of each paragraph with AI-generated questions. After reading, Marvista generates an explainable human-AI summary that combines both AI's processing of the text, the user's reading behavior, and user-generated data in the reading process. In contrast to prior work that offered (content-independent) interaction techniques or devices for reading, Marvista takes a human-AI collaborative approach that contributes text-specific guidance (content-aware) to support the entire reading process.

> *Reading ... requires much more complex psychological processes of strategic search, organization for remembering, use of natural units in problem solving, the discovery of rules and order, and the economical use of them.* —Gibson & Levin, The Psychology of Reading.

## 1 INTRODUCTION

Reading is a universally important activity that can also be challenging for many of us, as articulated in the above passage from [21]. Currently, 43 million US adults possess low literacy skills (< level 2) and even for people who can comprehend text, they might struggle to "truly understand and appreciate what the text is saying" [24, 62]. The growing consumption of online media might have

Authors' addresses: Xiang 'Anthony' Chen, UCLA HCI Research,, xac@ucla.edu; Chien-Sheng Wu, Salesforce Research,, wu.jason@salesforce.com; Lidiya Murakhovs'ka, Salesforce Research,, l.murakhovska@salesforce.com; Philippe Laban, Salesforce Research,, plaban@salesforce.com; Tong Niu, Salesforce Research,, tniu@salesforce.com; Wenhao Liu, Faire,, owenhaoliu@gmail.com; Caiming Xiong, Salesforce Research,, cxiong@salesforce.com.

exacerbated such challenges, as online media often comes in small bits that encourage "rapid, distracted sampling" while weakening our ability to focus, read, and think deeply [10], especially when the amount of information, such as long articles, goes well beyond most social media posts. As most text reading is taking place online [61], there are fertile opportunities to create digital tools that provide guidance to help readers more strategically navigate and comprehend text.

To support people's reading activities, past work has been studying and comparing how people read text on different platforms [46, 52], leading to a plethora of interaction techniques [13, 26, 35, 43, 44, 63] and devices [4, 28, 58, 73] that support active reading, *i.e.*, reading combined with "critical thinking and learning" by directly manipulating or annotating text. However, the problem is that, even provided with such techniques or devices, people performing active reading are still faced with challenges due to a lack of guidance with respect to the specific text a person is reading. For example, a reader might not know how to strategically navigate the text given their specific interest or limited time constraints. As people are increasingly overwhelmed by the amount of online information (*e.g.*, news articles), it is important to complement generic reading techniques with text-specific guidance to inform readers as they process text. In particular, the need for guidance when reading news articles (*e.g.*, developing awareness of news sources' credibility and the diversity of opinions) has been demonstrated and supported in prior work [33, 50, 54, 55]. Recent developments in natural language processing (NLP), such as text summarization models, are ready to serve for such purposes; yet it remains unknown how to design tools that harness such NLP models to support a reader.

To bridge this gap, we explore a human-AI collaborative approach via the design of Marvista—a news reading tool that employs a suite of NLP models[1] to understand the text and then to provide various guidances. The design goal of Marvista is to leverage AI-generated guidance to help people become aware of how they read an article and engage themselves throughout the entire reading process. Figure 1 summarizes the key human-AI collaborative features in Marvista spanning the entire reading process.

Before reading an article, Marvista helps a reader plan a high-level reading strategy using two filtering techniques: *(i)* The *time filter* allows the reader to specify how much time they would like to spend, based on which Marvista highlights the most summative subset of the text to read that stays within the user's time budget; and *(ii)* The *question filter* allows the reader to select from a list of AI-generated questions that can be answered by reading the article, based on which Marvista highlights the corresponding text that contains the answers.

During reading the article, Marvista modifies the display to let the user read the article one paragraph at a time (by "dimming" the other elements on the page). In this way, Marvista can also infer the user-perceived importance of each paragraph by recording how much the reader spent dwelling on that paragraph[2]. Besides the common active reading features of highlighting and note-taking, a Marvista user can also *reflect* on their understanding by answering an AI-generated question about a paragraph.

After reading the article, to help a reader consolidate their understanding, Marvista generates a *human-AI summary* that combines both AI's processing of the article's text as-is and the user's perceived importance of each paragraph. Specifically, the user can adjust how each factor should weigh in the summary, including their dwelling time per paragraph (implicit behavior) and user-generated data, *i.e.*, notes and highlighted text (explicit behavior). Further, such human-AI summary is explainable: the user can hover on each sentence in the summary to see how it can be attributed to the specific text in the article or the specific user-generated data that leads to certain text being included in the summary. Another design rationale of such an explanatory feature is providing users the opportunity to see "model hallucination" effects (if they exist), *i.e.*, an abstractive summary contains contents irrelevant to the article. If the user revisits the same article later, they can simply shortcut to this summary to remind themselves what they read (as well as how they read it).

---

[1]Hereafter interchangeably referred to as 'AI'.
[2]Currently, Marvista only records dwelling time on each paragraph, which would inevitably include time of non-reading activities.

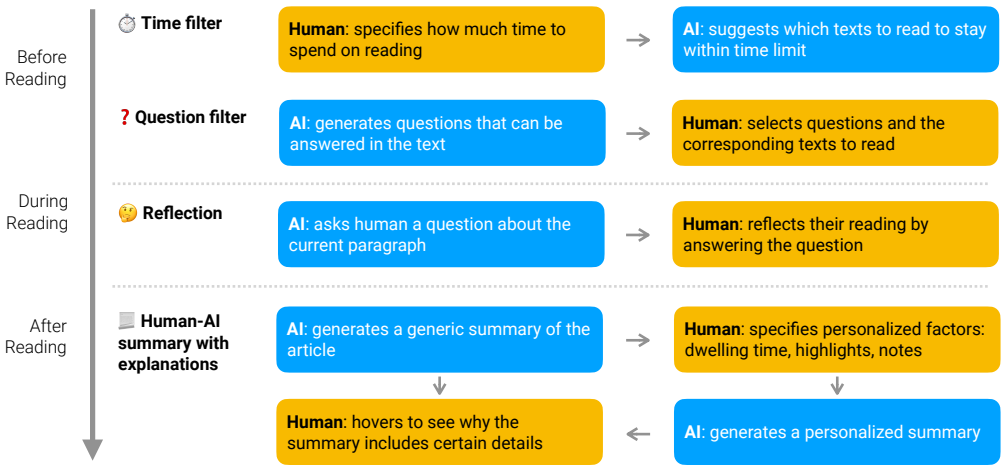

Fig. 1. We explore the design of Marvista—a human-AI collaborative tool that supports the entire process of reading an online article. Before reading: Marvista helps a user plan what to read by filtering text based on how much time one can spend (time filter) and what questions one is interested to find out (question filter); During reading: Marvista helps the user reflect on their understanding of each paragraph with an AI-generated question (reflection); After reading: Marvista generates a human-AI summary that combines both AI's processing of the text, the user's reading behavior, and user-generated data; further, hovering over a sentence explains which paragraph it is related to and why including its information in the summary.

In our user study of Marvista, the main research question is how users would interact with Marvista's AI-generated reading support. We conducted a study with 17 participants who open-endedly used Marvista to read science and technology news articles of their choice from the PopSci[3] website. Rather than measuring reading comprehension performance, our goal was to probe how people would interact with, think of, and benefit from Marvista's human-AI collaborative reading features. The most significant findings and insights are: *(i)* Participants used Marvista to better understand not only the text they read but also how they read the text (enhancing awareness); *(ii)* Marvista's different features are better fits for information-seeking tasks (where a reader needs support to stay engaged in the text) but not for casual reading (*e.g.*, for entertainment); *(iii)* Participants found Marvista different from how they normally read text; *(iv)* One challenge was to interpret Marvista's output with limited context; and *(v)* Instead of unconditionally trusting AI, multiple participants were able to identify issues with the model.

**The main contribution** of this work is the design of a human-AI collaborative reading news tool that leverages existing NLP models to provide end-to-end support for a user's reading process. We explore the role of AI as a "reading buddy" that provides text-specific guidance , from planning a custom reading strategy, to reflection upon reading a paragraph, and to generating a personalized, explainable human-AI summary after reading the article. Currently, we are in the process open-sourcing Marvista, which can soon be downloaded and installed as a Chrome extension. Now, Marvista supports over 5, 000 websites, *e.g.*, Popular Science, TechCrunch, and Medium.

---

[3]https://www.popsci.com

## 2  RELATED WORK

As a human-AI collaborative reading tool, Marvista is related to two schools of prior work: human-AI collaborative systems and tool support for reading.

### 2.1  Human-AI Collaborative Systems

The high-level principles of designing human-AI collaborative systems are discussed in early work. For example, Terveen characterizes two major approaches of humans collaborating with computers: *(i)* the human emulation approach where computers are built with "*human-like abilities* to enable them to *act like humans*" and *(ii)* the human complementary approach that leverages computers' "*unique abilities, to complement humans*" [64]. Horvitz lays out principles of designing mixed-initiative interaction where humans and agents take turns to jointly accomplish a task [29]. For example, whether an agent takes initiative should be determined by considering both the utility of the action against the uncertainty of the inferred user intents. A plethora of later work instantiated these principles into the design of specific interactive systems, which, for presentation purposes, we organize into the following non-mutually-exclusive categories.

*(i)* AI can take a user's partial input and continue with the rest of the task that would otherwise be manual and possibly tedious. For example, Chateau is a 3D sketching interface that observes what has been drawn and predicts a user's next drawings or provides suggestions to complete their current drawing [31]. Tsang *et al.* search for images similar to a target design to guide the creation of 3D curves and suggest relevant geometry based on users' input strokes [66]. In the domain of writing, numerous text entry techniques employ language models to autocomplete a user's typing [42] and recent developments in neural networks further enable 'smart compose'–AI that can autocomplete a user's partially written sentence [12] or improve writing skills [30].

*(ii)* AI can filter, transform, or perform analyses of a large amount of data so that humans can examine the data more efficiently to obtain more insights. Wang *et al.* interviews data scientists to learn their attitude towards AutoAI—AI that can potentially automate what data scientists do by analyzing data and creating preliminary models from data [70]. Similar to data scientists, medical professionals are also faced with challenging scales and complexity of data. Tschandl *et al.* compare how different types of AI-generated information can help or affect physicians' diagnosis behavior/outcome in a skin cancer domain [67]. Gu *et al.* found that pathologists benefited from AI-suggested regions to start the examination of a histological slide [22]. On the other hand, when AI algorithms cannot retrieve clinically meaningful past medical images, physicians' input can help filter and refine the results so that the examples are more relevant to a present case [9].

*(iii)* AI can offload low-level computational subtasks based on high-level goals specified by humans. For example, in Umetani *et al.*'s furniture design tool, as users edit a design, the tool indicates structures that are non-durable and/or non-stable, and further offers suggestions for solving these problems [68]. DreamSketch enables users to sketch their ideas (*e.g.*, the seating of a glider) while leaving structural components (*e.g.*, connectors between seats, handles, and wings) to be generated by an optimization module [32]. Willett *et al.* develop a tool for turning static images into animation where the automated algorithm is guided by a small amount of user input (*e.g.*, which objects to animate, the trajectory, and depth information) [71]. Forte is a tool where an optimization algorithm follows a user's rough sketch to generate structural designs and addresses their preference such as how much the design can deviate from the initial input [14].

Marvista's human-AI collaborative support for reading mostly resembles the second approach—AI processes text before, during, and after reading to help users read in a more efficient, focused, and informed manner; meanwhile, users' reading behavior and generated data helps AI provide more personalized support, *e.g.*, generating a human-AI summary that addresses a reader's interest (as dwelling time) on each paragraph.

## 2.2 Tool Support for Reading

Reading on digital screens (as opposed to on physical papers) was a relatively recent development that has been a result of increasingly democratized personal computers from laptops to tablet devices. To support screen reading, several studies have attempted to understand people's reading behavior from the physical to digital domains. O'Hara surveys literature to construct a typology of reading: how people read, what they do besides reading, and why they read [51]. Specifically, this typology incorporates Lunzer's taxonomy of four reading types (receptive, reflective, skim, and scan) [41], Anderson & Armbruster's categorization of reading support activities (underlining, note-taking, outlining, and networking) [1], and Robinson's SQ3R technique (survey, questions, receptive/recital/review) [60]. A follow-up study by O'Hara and Sellen compared the difference of reading on physical papers vs. digitally on a screen and summarized users' behavior and experience in terms of annotation, navigation, and layout [52]. This study was later revisited by Morris *et al.* with additions of new reading media (horizontal display with pen input and multiple tablets) [46].

All these studies gave rise to a series of interactive devices and techniques that support active reading on digital media. The term 'active reading' refers to "a broad set of cognitive skills and activities, such as thinking, learning, note-taking, annotations, searching and skimming, that enable an individual to achieve a deep level of comprehension of a document" [5]. One of the earliest active reading tools is XLibris—a hardware/software platform that supports active reading via three main features: a digital tablet for reading scanned images of papers, a "Reader's Notebook" to collect excerpts of text, and capabilities to search for similar parts of the document based on highlighting and annotations. The popularization of touch devices led to several designs, such as LiquidText—a tablet-based reader that employs a suite of multi-touch gestures to address specific problems in reading text on physical papers, such as organizing and synthesizing annotations and reading disparate parts of the document in parallel. Hinckley *et al.* present touch-based techniques for informal information gathering (*e.g.*, cropping a paragraph, collecting multiple pieces of images/text) that are designed to be lightweight and to not disrupt the flow of reading [27]. Mehta *et al.* study and develop tool support for literary critics to perform "close reading" of text where each reading connects to and builds on the previous ones and external resources are often drawn on to facilitate the interpretation [44]. Some research employs multiple types of devices for reading. Chen *et al.* design a multi-slate reading environment that mimics how people's reading activities often benefit from multiple spatially-distributed physical papers [13]. Matulic and Norrie incorporate pen input together with touch to support navigation and in-document search [43]. In addition to input, prior work also optimizes the display of text, such as anchoring salient objects during scrolling as landmarks to help a user navigates text [35].

In the meantime, some research has invented new device components to support active reading. Bianchi *et al.* design and build a physical reading aid for digital tablets that mimics a physical bookmark yet provides a suite of interaction techniques to support active reading [4]. Yoon *et al.* explores the sensing of grasp when using tablet devices to design techniques that enhance active reading for and across individuals [73].

While the above prior work does not defined a specific type of text for reading, others have focused on supporting reading news articles [33, 50, 54, 55] and research papers [20], *e.g.*, adding a tooltip that points a user to the definition of unfamiliar terms [2, 25]. While support for reading news articles mainly focused on sources' credibility [50] and diversity [33, 54, 55], our work generates guidance based on contents of news articles.

## 3 DESIGN & IMPLEMENTATION

The overarching goal of Marvista's design is to provide a reader with useful information about an article (acquired by AI preprocessing the text) throughout the entire process of reading that article. In this section, we describe Marvista's key features spanning three phases: before, during, and after reading an article. Rather than innovating new NLP techniques, our implementation leveraged existing models to realize Marvista's interactive features. Such NLP models' performances were reported in their original papers; thus we did not re-evaluate them individually.

### 3.1 Before Reading

Psychologist Eric Lunzer differentiated receptive *vs.* skim reading in the following way [41]: taking the receptive approach, one reads the text sequentially "as approximating listening behavior" whereas skimming requires a decision-making process about which parts of the text are worth reading. Unfortunately, at present there is little support for a reader to skim the text based on their specific constraints or interests. As a result, readers might attempt to receptively read most of the text only to find out they do not have time to finish it or that there are only specific parts that actually interest them. To fill this gap, Marvista employs NLP models to filter the text based on two user-defined criteria, which provides value in the active reading process by helping a user plan strategically what to read.

*3.1.1 Time filter.* As shown in Figure 2a, as a reader selects the time filter, Marvista first estimates the total amount of time required to read the entire article. Then the reader uses a slider to specify up to how much reading time they would like to spend and Marvista will select and highlight a subset of the text in blue (Figure 2b) that the system considers as must-read given the time constraint.

**Design rationale**: By allowing a reader to decide how much time they want to spend on reading (while still covering the most important content), the time filter intends to enable people to manage and plan their time to be spent on reading, *e.g.*, knowing how to skim the text when reading the article is part of a time-sensitive task or when the reader simply does not have the attention span for the whole article.

**Implementation**: Marvista estimates the total reading time $T$ of an article based on an averaged reading speed of 150 words per minute[4]. The reader's specified time limit $t$ is used to compute a compression rate ($t/T$), which then serves as a parameter for an extractive summarization model that ranks all input sentences [45]. We use one existing extractive summarization model utilizing a BERT model [17], where it uses the BERT model for text embeddings and K-Means clustering to identify sentences closest to the centroid for summary selection. More details about the model training, performance, and qualitative study are reported in the original paper [45] and its concurrent work [38]. Note that we select this method as a proof-of-concept solution and it can later be easily replaced with any other more advanced extractive summarization models. Our built extractive summarization solution will then return a set of sentences that are a subset of the source article whose total length is equal to or smaller than $t/T$ of the entire text's length.

*3.1.2 Question filter.* As shown in Figure 3a, a reader can also choose to filter the text by questions and Marvista will generate a list of questions that can be answered by reading specific parts of the article. The reader then selects any number of questions they are interested in and Marvista will highlight the corresponding sentences that answer these questions (Figure 3b). Once the reader starts reading the text, hovering a highlighted sentence brings up a tooltip (Figure 3c), showing the corresponding question that can be answered.

---

[4]We currently use an average reading speed, although the system can be easily extended to let the user specify their own reading speed. The 150 words per minute was a little lower than that in other studies (*e.g.*, 200-250, as reported in [65]) to consider that our current reading materials—science and technology articles—might be more difficult to read.

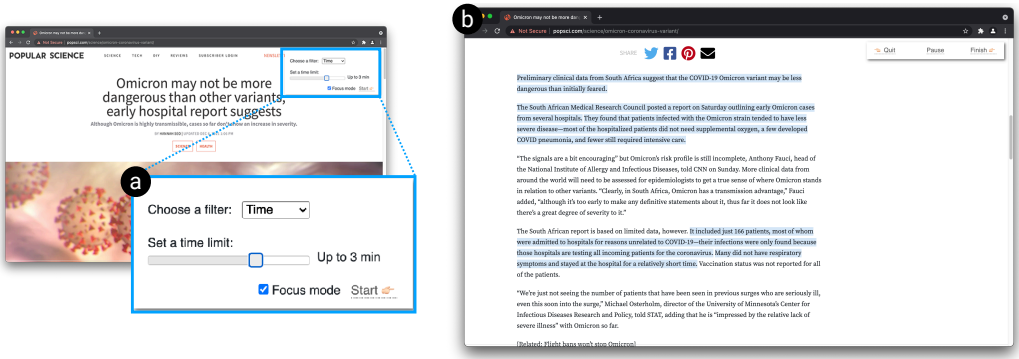

Fig. 2. Before reading an article, Marvista helps a user plan what to read by filtering text based on how much time one can spend, based on which Marvista highlights the most summative subset of the text to read that stays within the user's time budget (b).

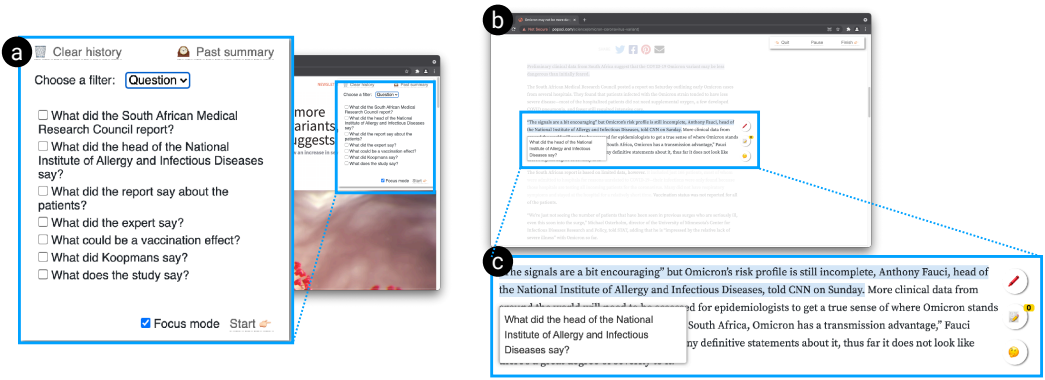

Fig. 3. Before reading an article, a user can also filter the text by selecting from a list of AI-generated questions (a) that can be answered by reading the article; Marvista then highlights the answers in the text and hovering over a sentence shows the corresponding question (bc).

**Design rationale**: The design of the question filter is informed by the SQ3R technique proposed by Robinson [60]. The SQ3R technique starts with a preliminary *survey* of the text, followed by the reader asking the kinds of *questions* that they think can be answered by reading the text, then continued with *receptive* reading interleaved with *recitals* of the main take-away points, and ended with *reviewing* what has been learned. Specifically, Marvista helps a reader with the "Q" step so that the reader can spend more of their time on the actual reading to answer their interested questions.

Both the time and question filters help a user decide what to read in an article. Complementary to existing heuristics (*e.g.*, the inverted pyramid of writing [56]) one can refer to as a general reading technique, Marvista's support provides further specific recommendation based on a user's constraints, interests, and the very contents of an article.

**Implementation**: Marvista employs an existing question generation (QG) model, MixQG [47], which is based on a large-scale pretrained model named T5 [59] and further fine-tuned on nine QA datasets. MixQG takes two parameters as input to generate a question: *(i)* the text that a question will be asked about and *(ii)* the text span that will serve as an answer to a desired question. Specifically, Marvista uses the text of the entire article as the first parameter. For the second parameter, Marvista populates the list of answers using the extractive summaries [45] of each paragraph. Paragraphs that are shorter than a pre-defined threshold (word count ≤ 32) are merged with the subsequent paragraph (if there is any). In this way, each generated question can often be answered by reading one specific paragraph (sometimes multiple ones if they are short).

The MixQG model we use has been already validated through human evaluation by prior work [34, 47]. In prior work, QG models (including MixQG models of varying sizes) were benchmarked on the "Quiz Design task" in which teachers working on creating reading comprehension quizzes are assisted by a QG model to design quiz questions. The results of the study indicate that the MixQG-Large model achieved the highest rate of accepted questions by the teachers. Prior work describes the three limitations of generated questions: *(i)* a lack of context, *(ii)* a lack of answerability, and *(iii)* a lack of fluency in the questions. These limitations are likely to appear in our interfaces as well, since we use the model as-is, which we discuss later in §7.5.4.

### 3.2 During Reading

Complementing existing active reading techniques that allow a user to generate contents (text annotations or notes), Marvista provides additional value by guiding a user to focus on one paragraph at a time and to reflect on the specific contents of a focused paragraph.

*3.2.1 Focused mode.* As shown in Figure 4, Marvista's focus mode displays one paragraph at a time while making the other paragraphs and UI elements translucent. A reader can use up/down arrow keys to go to the next or previous paragraph. The reader can also click 'Pause/Resume' (top right of the screen) to switch between the focused mode and the normal view of the article.

**Design rationale**: Although this feature does not employ any AI, we still include it in our design for two reasons: *(i)* To facilitate a more focused way of receptive reading [41, 60]. By seeing only one paragraph at a time and having to explicitly navigate to the next one, a reader is made more aware of what they are currently reading and less distracted by the other visual elements[5]. *(ii)* To estimate how much time a reader spends on each paragraph, which later will be necessary for generating a personalized human-AI summary (detailed in §3.3.2) that emphasizes each paragraph proportionally to the time spent by the particular reader. Such a human-AI summary is akin to the reader's 'book report' of the article because it represents which parts of the text the reader actually spends more time on.

**Implementation**: Marvista follows site-specific rules (*i.e.*, HTML selection queries) to extract HTML elements that correspond to paragraphs of an article. Reading time is tracked and accumulated for each paragraph whenever that paragraph is in focus, which includes both reading as well as time spent using the accompanying tools (described below).

*3.2.2 Reflection.* As shown in Figure 5, as a reader focuses on a paragraph, Marvista provides a highlighting feature and  tools to support active reading, including  note-taking and a novel AI-enabled feature—reflection. As a reader clicks the reflection button, Marvista generates a question about the current paragraph (*e.g.*, "What did the scientist say about Omicron?") for the reader to answer by reflecting on what they have read in the paragraph.

---

[5]One immediate trade-off of such focus mode is a loss of context, which we discuss later in §4.

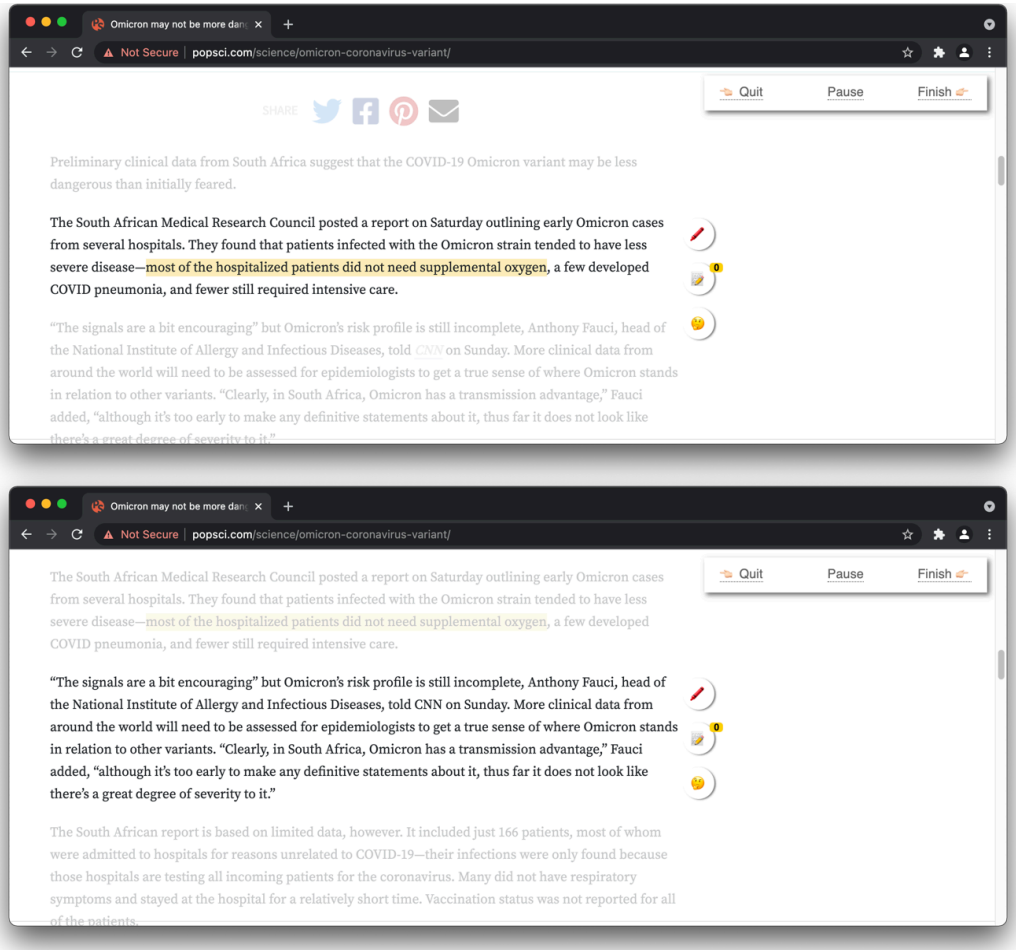

Fig. 4. During reading an article, Marvista's focus mode modifies the display to let the user read one paragraph at a time (by "dimming" the other paragraphs), using the ↑/↓ keys to navigate to the previous/next paragraph.

**Design rationale**: The design of this feature is informed by one of Lunzer's taxonomized reading types called *reflective reading* [41], which "involves interruptions by moments of reflective thought about the contents of the text". Currently we do not show the answer to the reflective question so that a reader cannot just directly reveal the answer without going through the reflection process. In contrast to prior tools that answers a user's questions [11], this feature flips the relationship and instead quizzes the user's understanding of the text they just read.

**Implementation**: We employ the same QG model as the one used in the question filter feature, except now we generate only one question using the extractive summary of the focused paragraph as the answer.

### 3.3 After Reading

Even after finishing the active reading of an article, Marvista continues to provide value to a user by reviewing what they have read via an AI-generated or a Human-AI summary, both of which provide explanation to further help the user understand how certain text is emphasized in the summary.

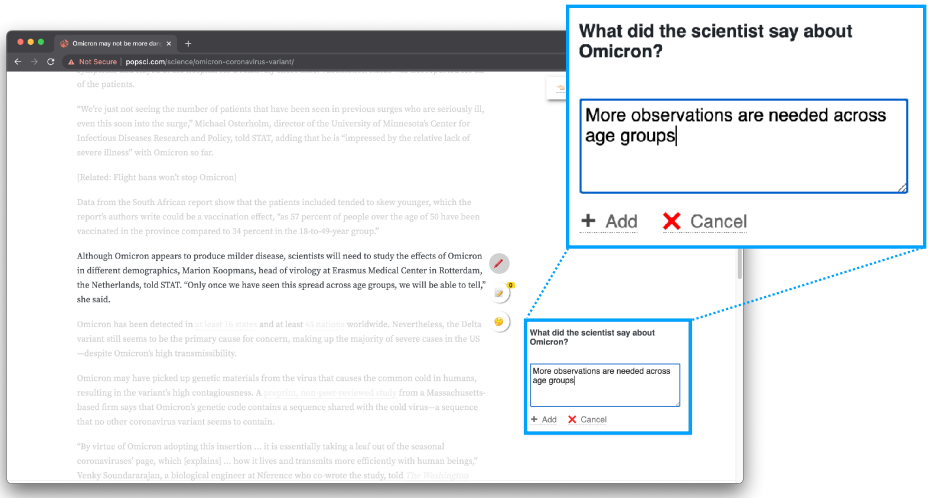

Fig. 5. During reading a paragraph, a Marvista user can reflect on their understanding by answering an AI-generated question about that paragraph.

*3.3.1 AI-generated summary.* As a reader clicks the "Finish" button, Marvista first generates a summary using an abstractive summarization model. While extractive methods summarize text using a subset of its sentences, an abstractive summarizer generates new (and often shorter) text that carries the most important information from the original text.

**Design rationale**: This feature allows AI to assist the reader with the "review" step in Robinson's SQ3R technique [60]. We also envision that such a summary would help the reader recall the important information when they revisit the article later. Currently, Marvista saves the most recent reading session and, instead of re-reading the article, a returning reader can choose to directly go to the summary page (by clicking 'Past Summary' as shown in Figure 3a). While currently Marvista only provides a summary at the end of reading, it is possible to also prompt the user with a summative question to encourage them to reflect on the entire article.

**Implementation**: Marvista uses the standard Transformers [69] encoder-decoder architecture to generate an abstractive summary as the encoder, uses RoBERTa [39] as the decoder, and then finedtunes the summarizer with a cross-entropy loss on the CNN/Daily Mail [48] summarization dataset, which is similar in nature to the science and technology news articles in our reading tasks. We use Longformer [3] because it can take eight times more tokens as input than the commonly used BERT encoder, which handles long articles better. The model[6] achieves ROUGE2 F-measure 13.21 on CNN/DailyMail dataset, which is by no means a state-of-the-art model but it can produce reasonable and high quality summarization results as a proof-of-concept solution. With the recent success of OpenAI's InstructGPT and ChatGPT [53], we plan to switch our abstractive summarization backbone in future work.

---

[6]https://huggingface.co/patrickvonplaten/longformer2roberta-cnn_dailymail-fp16

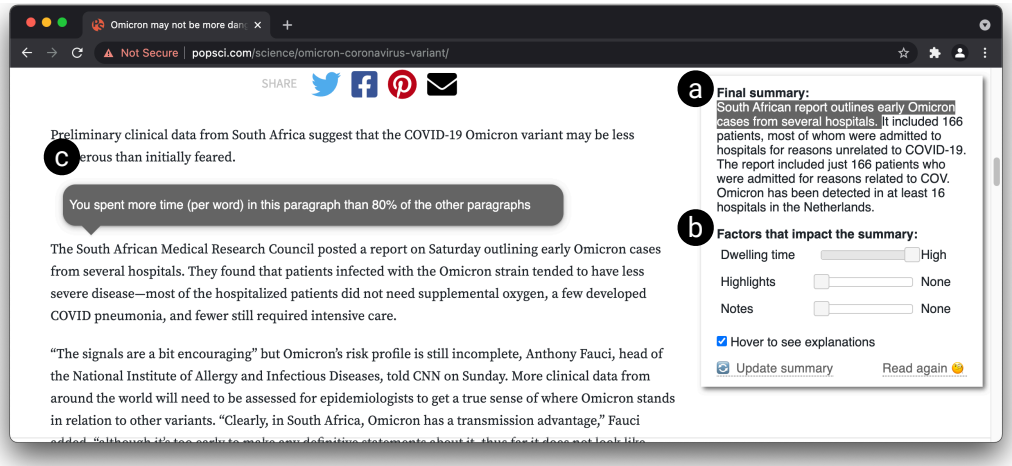

Fig. 6. After reading the article, Marvista generates a human-AI summary (a) that combines both AI's processing of the text as-is and the three personalized factors (b): the user's dwelling time on each paragraph, user-generated notes, and highlighted text. Further, the summary is explainable: the user can hover on each sentence in the summary (a) to see how it can be attributed to the specific text in the article or the specific user-generated data that leads to certain text being included in the summary (c).

*3.3.2 Human-AI summary.* Beyond the aforementioned one-size-fits-all summary, Marvista further allows the reader to personalize the AI-generated summary based on two sources of user-generated data from the reading process:

- *Implicitly*-generated user data, for which we use the reader's dwelling time on each paragraph to weigh its importance. To factor in different lengths across paragraphs, we calculate reading speed for each paragraph (number of words divided by dwelling time[7]): the more time a reader spends per word in a paragraph, the more likely that paragraph will contribute to the generated summary;
- *Explicitly*-generated user data, which includes the reader's notes and highlighted sentences as additional contents to the original text. The combined text is then fed to the model to produce an abstractive summary. Admittedly, sometimes users might highlight irrelevant or unimportant sentences—our future work will experiment with a feature to detect these 'outliers' and let the user decide whether they want to leave out such data.

A shown in Figure 6b, for each type of user-generated data, the reader can choose a level of impact on the summary: none, low, or high.

**Design rationale**: Dwelling time at the word level has been used in prior work for training a model to estimate a reader's interest on unseen text [60]. Analogously, Marvista uses dwelling time at the paragraph level to measure a reader's interest and enables them to see their interested paragraphs emphasized in the summary. Annotation (*e.g.*, highlighting and note-taking) is an integral part of the active reading process [51], yet such user-generated data is often scattered across the documents. Marvista's human-AI summary feature allows the reader to see their notes and highlighted sentences integrated into the summary. This feature is similar to prior work that incorporates readers' manual filtering or implicit feedback (*e.g.*, gaze) to personalize a document summary [6].

---

[7]Dwelling time is only available when the reader used the focused mode; otherwise, the control of this factor will be disabled.

**Implementation**: To incorporate dwelling time into the summary, we normalize the aforementioned reading speed and map it to weights for each paragraph. We then "hijack" the abstractive model with these weights by element-wise multiplying them to the latent encoder representations of each paragraph before they are fed to the decoder. Note that in this way we can manipulate output summary by adjusting its encoder's latent space implicitly without the need to collect personalized summaries as training data.

To determine the range of weight values, we conducted an experiment using a small corpus of text from three articles. Given the paragraphs of an article, we gradually increased the weight of one paragraph $P$ from 0 with a step of 0.1 while fixing the others' weights as 1. We then computed the ROUGE score [36]—a similarity metric—between the generated summary and $P$, and found that the score increased/decreased with the weight but flattened below 0.6 and above 1.4. Thus we use [0.6, 1.4] as the range, *i.e.*, an unread paragraph will be weighed as 0.6 and the paragraph with the highest time per word as 1.4. Currently, we use an exponential mapping from time-per-word to weight so that paragraphs with greater dwelling time would stand out more and vice versa.

To incorporate notes and user-highlighted sentences, we simply insert each note or highlighted sentence as new paragraphs immediately after the original paragraph wherein they were generated. Then such user-generated data is treated identically as the other paragraphs by the summarization model. The three levels of impact are mapped to specific weight values for the paragraphs of highlights or notes: *none*=0.6, *low*=1.0, and *high*=1.4.

*3.3.3 Explanation of summary.* As shown in Figure 6c, as a summary is generated, the reader can check the box of 'hover to see explanation' feature (Figure 6b). Then, as they hover over a sentence in the summary, Marvista locates the source of text (or user-generated data) and describes how it is correlated with the summary sentence in the explanation bubble.

**Design rationale**: The goal of such explanation feature is two-fold: *(i)* to create a feedback loop so that the reader can see that the summary indeed incorporates user-generated data as they specify; and *(ii)* to serve for indexing purposes so that the reader can use the generated summary as a "portal" to access specific reading history, notes, and highlighted text.

**Implementation**: We employ a post hoc similarity-based approach to generate such explanations. Given a hovered sentence $S$ in the summary, we compute the ROUGE score [36] between that sentence and each one of the original paragraphs, user-generated notes, and user-highlighted sentences[8]. If the highest score is between $S$ and an original paragraph, we first check its dwelling time (per word): if it is above the average, we report in the explanation "You spent more time (per word) in this paragraph than $X\%$ of the other paragraphs"; otherwise, we show in the explanation bubble "This paragraph is the most related." If the highest score is between $S$ and a note or a highlighted sentence, the explanation bubble shows "You wrote a related note here: [the note]" or "You highlighted a related sentence here."

Our generated explanation uses ROUGE score, which determines how an explanation (*e.g.*, a paragraph in the article) is relevant to a sentence of the summary. Recent work [37] has shown that in a reference-based context—in which a known reference is the gold standard target—ROUGE remains a strong metric to evaluate summary relevance. We believe our explanation matching process resembles the reference-based evaluation setting as they both involve rating text similarities, and the recent work therefore validates the use of the ROUGE metric in our implementation.

---

[8]Unless the impact level is set to *none* for the latter two factors.

## 4 USER STUDY: PROBING HOW USERS OPEN-ENDEDLY INTERACT WITH MARVISTA

While it is possible to compare our approach with traditional tools on users' reading comprehension performance, such an evaluation might be premature without first understanding how readers would use human-AI collaborative tools like Marvista. Thus, in this paper, we chose to focus on conducting an open-ended study to probe users' reading behavior while interacting with Marvista, while leaving task-specific reading performance evaluation as future work.

### 4.1 Participants, Tasks, & Procedure

We recruited 17 participants (nine males and eight females, aged between 20 and 39) from a local university. There were six undergraduate, three master, and eight doctoral students, majoring in Electrical & Computer Engineering and Mechanical Engineering. In a pre-study reading behavior survey, we asked participants how often they read (when it was their choice): seven reported daily, eight at least once per week, and two every few months. When asked how much time they spent each time they read (when it was their choice), eight reported 15 to 30 minutes, six reported 30 minutes to an hour, and three reported over an hour. When asked their types of reading material, 14 included articles on a website. When asked how much they preferred reading from a computer screen (1: least preferred ↔ 5: most preferred), 14 reported equal to or higher than neutral preference. When asked what motivated them to read (multiple choices), 12 chose "required in my study or work", 11 chose "need information for myself", and nine chose "enjoyment".

The study was approved by our Institutional Review Boards (IRB) and conducted in person[9]. We first went through the informed consent process with each participant, including an explanation of the kinds of data that would be collected, i.e., audio and screen recordings and their usage log of interacting with our tool. Specifically, the log consisted of mouse/keyboard input events and user-generated data (e.g., their answers to reflective questions), which would be stored in the browser's cookie and deleted after we processed the raw data after each study session. Once the briefing was completed, we started with each participant watching a tutorial video to learn how to use Marvista, which took about five minutes. Next, they would spend about 25 minutes using Marvista to freely explore and read any articles on the Pop Science website. We chose this website because science and technology news articles have been commonly used for studying reading behaviors in past research [72]. To keep the study as open-ended as possible, we did not control which articles participants read. While participants might have chosen to read different articles, all the articles were about latest news in science and technology and were highly similar in the writing style. Participants used a Google Chrome browser to read articles that were displayed on a 13" laptop computer. All the NLP models ran on a separate networked server. Finally, we conducted a 30-minute interview to gather participants' feedback on each of Marvista's key features. Each study session lasted for about an hour and each participant was compensated with a $25 Amazon gift card for their time.

### 4.2 Measurement & Analysis

We recorded the screen and audio of the whole process. Integrated with Marvista, we logged participants' tool usage data, *e.g.*, which articles they read, how much time they spent on each article, and how often they interacted with each feature (detailed in §5). We analyzed this quantitative data in two folds: an overall analysis that computes the distribution of several key metrics and a phase-specific analysis that further looks into how participants interacted with Marvista's features before, during, and after reading an article.

---

[9]We conducted this study in-person in part due to the need of ensuring stable and fast network connection that was necessary for our tool to communicate with the back-end server. At the time of the study, our institution was already open for in-person attendance and we followed regulations throughout the studies such as in-door masking requirements.

During the interview, to gain further insight behind how participants used Marvista, we asked them to rate each of Marvista's key features (shown later in Table 1) based on the following two questions: *(i)* Does this feature provide value to your reading experience (1: not at all ⟵⟶ 7: highly valuable)? If so, what are the use cases? *(ii)* Does this feature require effort in addition to reading the article (1: not at all ⟵⟶ 7: high effort)? If so, describe the effort. For the qualitative data, we employed the thematic analysis approach [7]: the first author transcribed participants' responses to formulate the initial codes, which were then reviewed by two other authors. Then, disagreements were resolved via discussion between the authors.

Below, we report two sets of findings: quantitative data that characterize participants' reading behavior using Marvista (§5) and their qualitative feedback on Marvista (§6).

## 5 FINDINGS: READING BEHAVIOR USING MARVISTA

As shown in the histograms[10] in Figure 7, participants read an average of 3.8 articles (Figure 7a) and the average length (word count) of the articles was 928 (Figure 7b). On average, each article was read 1.9 times (each time is called a session): as shown in Figure 7c, most participants read an article up to two times, although a small number went as high as six or more times. Participants spent on average five minutes four seconds on each article (or two minutes 39 seconds per session): as shown in Figure 7de, the time-spending patterns per article and per session exhibit a 'long tail' with the majority on the lower end. When using the focus mode, we tracked which paragraph was actually being read and calculated the percentage of an article actually being read by a participant, which was on average 64.0% (Figure 7f): while a majority did read the entire article while using the focus mode, quite a number of sessions did not involve any reading at all. Upon further analyses, we found that amongst the 29 such no-actual-reading sessions, participants skipped to the summary 12 times by hitting the 'Finish' button, another 12 times they hit the 'Quit' button (which would lead them to the before-reading state of the tool), and in the remaining five times the page was redirected elsewhere (*e.g.*, going back to the main page or opening a hyperlink).

### 5.1 Usage of Before-Reading Features

Before a reading session of an article, participants filtered the text by time 41.1% of the time, by question 35.5% of the time, and without using either filter 23.4% of the time. For each participant, we calculated their average number of time- and question- filter usages per article and show the distributions in Figure 8ab. When using the question filter, participants selected an average 39.2% of the questions generated by Marvista (by default no questions were selected). When using the time filter, the reading time set by the participants was on average 57.6% of the estimated time of reading the entire article, which is not surprising considering the default setting was 50.0%.

To further understand how participants used the time filter, for each reading session, we compare the reading time set by a participant and the actual time spent in that session. Figure 8c showed the results: we can see that overall participants tended to spend less time than they had set using the time filter. We also compare time spent on sessions using the time filter *vs.* using no filter. On average, participants spent two minutes 30 seconds per 1000 words when using the time filter *vs.* four minutes with no filter. In other words, participants spent 37.5% less time on articles when using the time filter.

---

[10]In each of the histograms the Y-axis is the count and the X-axis is a specific measurement.

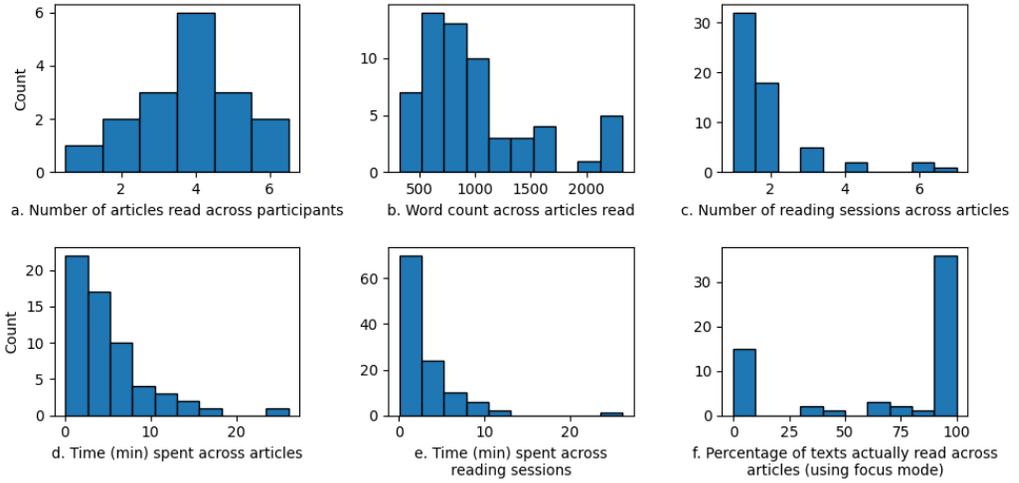

Fig. 7. Overall reading behaviors using Marvista

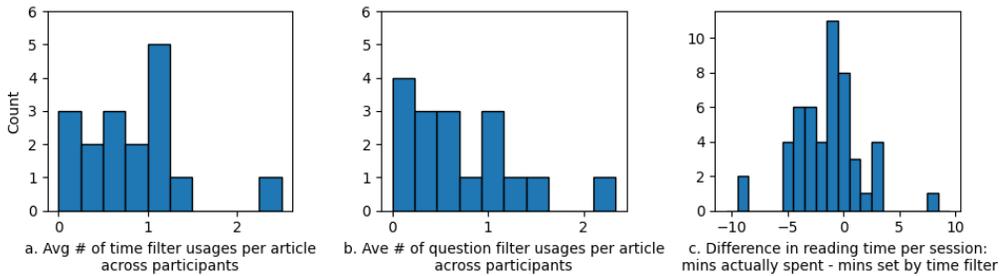

Fig. 8. Before reading: how often participants used time (a) and question filters (b) and how much time they actually spent after setting a time filter (c).

## 5.2 Usage of During-Reading Features

Participants used the focus mode 92.0% of the time. We hypothesize that the focus mode being the default setting might have contributed to such high frequency of usage but will need to verify that with further studies (*e.g.*, using an A/B testing setup). Overall, participants did not use the active reading features very often. The average number of usages per article were 1.2 for highlight, 0.9 for note, and 1.2 for reflection. Figure 9 shows more details about participants' distribution in their usage of the three features (per article). In particular, unlike highlight and note, there were two avid users of the reflection feature with a significantly higher number of usages (11 and 17 per article).

## 5.3 Usage of After-Reading Features

By design, participants were always shown an AI-generated summary upon finishing an article; in comparison, about two thirds (67.7%) continued to use Marvista to generate a human-AI summary. The average number of human-AI summaries generated per finished reading session was 1.0 and the distribution across participants is shown in Figure 10a. When choosing which one or more factors to impact the human-AI summary, participants overall seemed to prefer emphasizing dwelling time and highlights, choosing per-paragraph dwelling time in 64.3% of the cases (38.1% low impact and 26.2% high impact), choosing highlights in 64.3% of the cases (36.9% low impact and 27.4% high impact), and choosing notes in 26.2% of the cases (15.5% low impact and 10.7% high impact). Further, upon finishing a reading session, 59.3% of the time participants would seek an explanation of the generated summary (averaged 2.3 explanations per finished reading session). The distribution across participants is shown in Figure 10b.

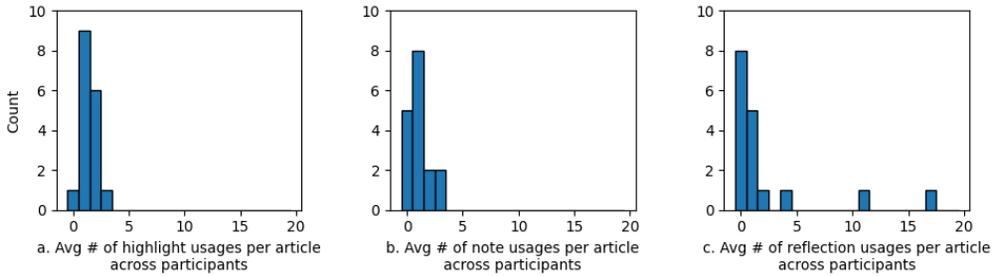

Fig. 9. During reading: participants did not use the three features (highlighting, note-taking, and reflection) very often.

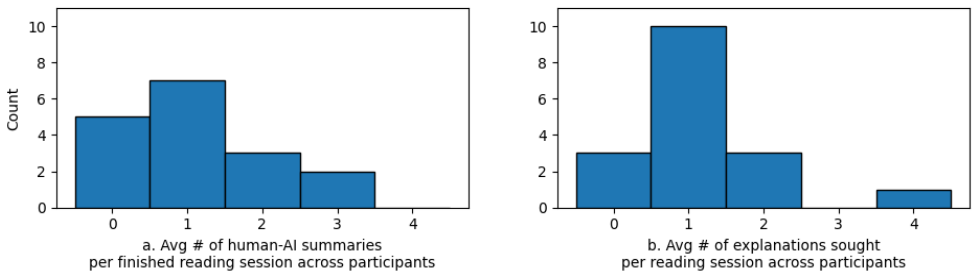

Fig. 10. After reading: how often participants generated a human-AI summary (a) and how often they sought an explanation of the generated summary (b).

### 5.4 Summary of Reading Behavior While Using Marvista

The above findings mainly indicate how likely users would interact with each of Marvista's features. Before reading, in over three quarters of the cases, participants made use of the time and question filters. During reading, particpants rarely opted out of the default focus mode but only scarcely used the active reading features (including AI-generated questions for reflection). After reading, in addition to the default AI-generated summary, in about two thirds of the time, a participant would generate a human-AI summary (with more emphases on dwelling time and highlight). In over half the time, participants would hover over the generated summary to see an explanation.

### 6 FINDINGS: USER FEEDBACKS OF MARVISTA

Table 1 visualizes participants ratings on each Marvista feature's value *vs.* effort. After each rating, we asked follow-up questions, including use cases for which each feature would provide value, the kinds of effort required, and other feedbacks and comments. We discuss the qualitative findings below, organized by features, and summarize recurring themes.

| | Does this feature provide value? (1: not at all; 7: highly valuable) | | | | | | | Does this feature require effort? (1: not at all; 7: high effort) | | | | | | |
|---|---|---|---|---|---|---|---|---|---|---|---|---|---|---|
| | 1 | 2 | 3 | 4 | 5 | 6 | 7 | 1 | 2 | 3 | 4 | 5 | 6 | 7 |
| Time filter | 0 | 1 | 1 | 3 | 2 | 3 | 7 | 5 | 4 | 5 | 2 | 0 | 1 | 0 |
| Question filter | 0 | 0 | 2 | 3 | 5 | 0 | 7 | 5 | 4 | 3 | 1 | 3 | 1 | 0 |
| Focus mode | 0 | 0 | 0 | 2 | 3 | 6 | 6 | 13 | 1 | 0 | 2 | 1 | 0 | 0 |
| Reflection | 1 | 1 | 0 | 2 | 3 | 5 | 5 | 6 | 4 | 1 | 2 | 3 | 1 | 0 |
| AI-generated summary | 0 | 0 | 0 | 2 | 5 | 4 | 6 | 11 | 4 | 1 | 0 | 1 | 0 | 0 |
| Human-AI summary | 0 | 0 | 0 | 2 | 4 | 3 | 8 | 6 | 8 | 0 | 0 | 2 | 1 | 0 |
| Explanation of summary | 0 | 0 | 0 | 2 | 3 | 5 | 7 | 7 | 5 | 3 | 1 | 0 | 1 | 0 |

Table 1. Participants' rating on whether Marvista provides value / requires effort in addition to reading an article.

## 6.1 Feedbacks to Individual Features

*6.1.1 Time filter.* Most participants were able to relate to the need of using the time filter. For example, P4, P5, P9, and P11 mentioned the use case of this feature for reading articles as part of work or study assignments, which requires reading efficiency. P6, P8, and P15 also commented that, by setting a certain amount of time and seeing the corresponding parts of the text, they felt they could better focus on reading. P12 and P13 mentioned the time filter's usefulness for skipping parts of long articles:

> "... useful for longer articles where the reader has an interest but not enough time to go through the whole article" (P12)

Some participants recognized that such a feature was not always useful, *e.g.*, P5 would not use the time filter when reading for leisure and P15 felt having to read fast was not the usual case for him. Some others pointed out the difficulty of deciding how much time one should choose (P1) and that, for novice users, it might not be clear what it means to read, say, three minutes of a five-minute article (P17). Another issue raised by P3, P8, and P12 was that AI-selected sentences sometimes were out of context and could not be fully understood as-is, which is a limitation of the extractive summarization model, as discussed later in §7.5.4.

As a response to this feature, we also saw two contrastive perceptions of time in reading. P9 and P11 mentioned the pressure they felt to finish the article within the set time even though Marvista did not actually enforce the reading time, while P2 and P15 felt the opposite and would like to see their reading time precisely tracked as they read the article. Such a disparity suggests the need of customized granularity when setting the time filter. For example, in lieu of exact time that might add pressure to some readers, Marvista can provide coarse modes: *e.g.*, quick, normal, or deep read; contrastively, Marvista can optionally add a timer during reading for those who prefer to track the time spent.

*6.1.2 Question filter.* Multiple participants (P3, P4, P9, P10, and P15) mentioned that the generated questions were useful and equivalent to the main ideas and key points of the article for them to preview before reading:

> "Helps check understanding. Make[s] sure you're getting main points. Shows you what to look out for in the article." (P10)

P7, P11 and P16 also pointed out that this feature was time-saving because a reader could spend their time only on questions that interest them. Further, some participants considered such questions a way to get them thinking about the article:

> *"When the article is too long or complicated, I might be a bit confused or unsure about my expectation after reading it. A list of questions give me some insights (and) reminders"* (P2)

> *"It primes my mind for the important parts of the article. Gets me thinking right away. It also leads me through the hover feature to the areas of the text where I can find the answer. I also found it efficient because I can isolate the parts of the text and just read the parts associated with the answer."* (P8)

As indicated in the effort rating (Table 1), the main issue brought up by participants was the difficulty/effort of choosing questions because *(i)* a reader unfamiliar with the article's topic might find it hard to fully understand the questions (P1), *(ii)* some questions needed more context (*e.g.*, "What did the scientists find out?") that was not available before reading (P3 and P17), and *(iii)* the process of reading through the questions and determining which ones to include could be effortful (P2, P5, P10, P11, and P15).

*6.1.3 Focus mode.* Focus mode is the only feature that did not directly rely on AI, yet it was universally well-received by all participants (all ratings on its value were neural or above).

> *"I think it really helped me focus on the specific paragraphs rather than being distracted by other elements."* (P14)

Participants pointed out the value of "*visually focusing*" (P1) on the current paragraph to avoid being distracted by other elements on the page (P4 and P14), which they considered helpful given that articles could be "*less organized*" (P9) and contain "*a lot of information*" (P7). Focus mode makes reading "*less overwhelming*" (P5) and "*... helped to break down the content and make it feel easier to move through.*" (P8)

Participants also pointed out this feature's usefulness when reading in "*a crowded room*" (P8) or "*a distracting environment*" (P15). A few also commented that they felt more aware of the paragraph they were reading (P3) and could go into more depth in the focus mode (P10).

While almost all participants perceived no extra effort in using this feature, two mentioned the trade-off of not being able to read the focused paragraph with more context, *e.g.*, looking back to already-read text to find relevant information (P16) or looking forward to skip to a certain paragraph. Another issue was adapting to such a different way of paragraph-by-paragraph reading: participants commented that using the arrow keys felt less natural than just normally using gaze (P2), sometimes causing one to forget to use the keys and just to habitually read the next paragraph without pressing the ↓ key to place it in focus (P5).

*6.1.4 Reflection.* Seven participants considered this feature useful for helping them check their understanding after reading a paragraph, which they "*may have overlooked when reading quickly or getting distracted*" (P8). P4 and P7 mentioned that they would use reflection to help them review or re-read a paragraph:

> *"... to force me to read over a paragraph again, in a more focused way. For example, in the articles I was confused about certain paragraphs, but had a hard time pinpointing my confusion. Reflecton tool gave me a more specific point to look for when re-reading."* (P4)

Meanwhile, three participants stated that they would not use this feature due to a lack of such needs (P17) or when reading "*for entertainment or quick info searching ... [and not wanting to] spend time answering questions*" (P2) or when preferring reading without interruptions (P12).

Multiple participants acknowledged the extra effort in re-reading the paragraph, thinking about and typing in the answers to the reflection question but all considered such effort "*worth it*" (P5 and P10) and "*valuable for improving a user's understanding of the paragraph*" (P13).

One noticeable issue, which was related to the question filter, was the quality of the generated reflection question. P10 commented that some questions focused on a factual detail of a sentence rather than capturing the overall idea of the paragraph and P16 felt some questions did not seem important with respect to the paragraphs. This is a known issue in question generation research in NLP, which we discuss in more technical details in §7.5.4.

*6.1.5 AI-generated summary.* Participants split on how they would make use of this feature. A number of participants (P1, P2, P5, P10, P14, and P15) considered AI's summary as an alternative solution to having to read the entire article thoroughly—

> *"Useful for someone who skims the article and would prefer to see a summary in lieu of a read-through."* (P5)

For some others (P3, P4, P5, P7, and P12), instead of skipping parts of the article, they would use AI's summary to compare with and check their own understanding of the article:

> *"As a 'check your understanding', for me. I like to have verification of my understanding of content but don't like answering questions. It's helpful to read a summary and compare to my understanding"* (P4)

The main issue, which was pointed out by seven participants, was the occasional inaccuracy of the model in capturing the gist of the article, which is a known issue in most summarization models. For example, P10 mentioned the summary was "sometimes too trivial" and did not "pick out what I thought to be the most important parts of the article". However, for these participants, such observed inaccuracy did not discount their perceived usefulness of AI's summary—they were cognizant that an AI-generated summary's usefulness came with certain limits and would choose to rely on their own when the AI was inaccurate—

> *"After reading I already had a basic understanding of this article, if the summary is accurate it might help me find something I miss; if not, it is still OK to me"* (P16)

*6.1.6 Human-AI summary.* Many participants liked how a human-AI summary further made the AI-generated summary "*more personalized*" (P3), "*useful to reflect readers' interest*" (P2), "caters" to what the reader thinks (P8 and P12), "*allows a user to control the AI summary*" (P17), and could incorporate each user's desired focus (P13)—

> *When adding comment into consideration, I feel like the summary is more personalized for me.* (P3)

Similar to their feedback on AI-generated summary, a subset of participants would use the human-AI summary to reflect on their reading of the article, *e.g.*, using the dwelling time factor to reflect on where they had spent most time on (P3 and P4) or "*to collect points that you read the most, but forgot to make a note of*" (P11).

> *"Having the dwelling and highlighting feature was pretty useful for me to quickly understand what was mentioned in the whole article as I may have forgotten what was written after reading everything."* (P9)

Participants also mentioned the possibility of influencing the human-AI summary during the reading process by intentionally highlighting or taking notes of contents that one would prefer being incorporated in the summary (P5, P6 and P14).

*6.1.7 Explanation of summary.* Participants provided diverse feedbacks on the value of summary explanations for addressing their curiosity (P3) and trust (P8) of the summarization model as well as contextualizing (P10) or deepening (P11) one's understanding of the article.

Some participants used this feature as a bridge between the short summary to the recalling of more details in the article (P1 and P6) and to connect the summary to the "right sources in the article" (P2, P4, and P5).

> "... useful if you don't remember something mentioned in the summary, so you could go back to the source and reread what you missed." (P5)

Similar to and building upon the human-AI summary feature, we again saw the recurring theme of multiple participants (P4, P5, P6, P7, P9, P10, and P17) using explanations to reflect on one's reading process, *e.g.*, which paragraph one spent more time on (P7 and P10) or to review important paragraphs (P6 and P15) or those that one might have missed (P5 and P9).

A few participants preferred to use this feature more selectively, *e.g.*, when "*looking for verification*" (P8) of the generated summary or only when one's thoughts did not align with the generated summary (P12)—

> "I think it's valuable only when I want to go back and check if the summary is generated correctly according to what I want." (P13)

In terms of efforts, participants did notice the extra work of going back and forth between the summary and the explanations inserted into the original article (P5, P12, and P13). Quite a few participants almost missed this feature because it was not enabled by default and required a reader to check a box (Figure 6). Two participants (P2 and P5) suggested turning explanations on by default.

## 6.2 Recurring Themes

**Participants used Marvista not only to better understand the text they read but also to better understand how they read the text**. Participants valued how the time and question filters could make reading more efficient, how focus mode could help a reader concentrate on the current paragraph, and how the summaries add to their understanding of the entire article. In the mean time, participants also reported using Marvista to better understand their own reading of the text, *e.g.*, how they used the reflection tool to check their understanding of the paragraph. Even for the three after-reading features that were intended to help readers better understand the text, many participants still mentioned using those features to reflect on how they had read the article (*e.g.*, to reflect where they had spent most time on). This suggests that summarization models can be beneficial and should be used for purposes beyond just summarizing text, just like how question generation can be beneficial for both planning what to read (*e.g.*, question filter) and reflecting on how one has read certain text (*e.g.*, reflection tool).

**Participants found Marvista changed how they normally read**. Many of Marvista's features required participants to employ a quite different reading style than what they were used to, *e.g.*, estimating how much time they want to spend before reading, reading one paragraph (only) at a time, pausing reading to answer reflective questions, and iteratively generating summaries after reading. The combination of all these features might significantly disrupt how people normally read. Even if a user chooses to read an article as-is (*i.e.*, without using any before- or during- reading features), they might pay more attention to how they read the text, knowing that such information can be used later to generate a human-AI summary. Participants' feedbacks suggest that they were still trying to learn such a new process and we should expect their reading behavior with Marvista to evolve over time and the perceived value and effort to change accordingly.

**Marvista's different features are better fits for some reading scenarios but not for the others**. Marvista is not a panacea for all reading scenarios. For example, participants considered the filters and generated summaries useful when reading to learn or acquire information but not when reading for entertainment, *e.g.*, feeling pressured after setting a time limit. Similarly, reflecting on a paragraph was considered useful only when one intended to read the article thoroughly but not for a quick read to search for specific information.

**One challenge is a lack of context, not the extra effort**. Although participants noticed extra efforts in using several features, such efforts never seemed to have prevented their usage of Marvista. The challenge that often mattered seemed to be a lack of context, *e.g.*, one challenge of the focus mode is trading off context for the focus of the current paragraph. Perhaps another more noticeable challenge is to understand an AI-generated question without context before reading or to understand isolated sentences AI extracts based on the filters. To mitigate this issue, one solution is to decrease the granularity, *e.g.*, AI suggesting what paragraphs to read instead of sentences when using the time or question filter. Further, explanations could potentially fill the gap of context, as demonstrated in the feature of summary explanations and reported by participants about how they used such explanations to connect the summary to contextual information in the article.

**Instead of unconditionally trusting AI, multiple participants were able to identify issues with the model**. Eight participants mentioned that they noticed a case where the model's output (either a generated summary or question) seemed inaccurate, mainly because of including trivial details while missing what participants thought was important. Such awareness of AI's issues also suggests that these participants likely compared AI's output to their own thinking rather than unconditionally trusting AI even when AI might be wrong (a behavior called "over-reliance on AI" [8]). **Over-reliance on tools like Marvista can also happen when users' understanding of the article depends more on AI's output than their own reading and thinking.** Indeed, as pointed out by one participant, "*[for] a developing reader it might make them lazy or dependent on it for finding the important parts of an article*" (P8). To further look into possible over-reliance on Marvista, below we report a follow-up study.

## 7 LIMITATIONS, DISCUSSIONS, AND FUTURE WORK

### 7.1 Further Evaluations of Reading Performance Using Marvista

Currently, we chose to focus on conducting an open-ended study to probe users' reading behavior while interacting with Marvista. Our findings suggest that participants considered Marvista more valuable when reading involves specific information-seeking tasks, *e.g.*, gathering relevant data for writing a report. On the other hand, Marvista seems unnecessary for casual reading, *i.e.*, reading for fun or entertainment. Based on such findings, our future work will conduct controlled comparison studies that instrument appropriate reading tasks (*e.g.*, reading articles of a specific topic to write a report), metrics, and measurements to evaluate reading performance using Marvista *vs.* a traditional tool. Such evaluation will include both low-level quiz-like questions after reading an article as well as high-level assessment, such as grading the quality of a report based on reading multiple articles.

In our future evaluations, we also plan to include longer articles (*e.g.*, those from Longreads[11]) into the reading tasks to test our hypothesis that Marvista might be more valuable to readers in navigating such larger amount of text. Note that these articles also go beyond the news domain, thus requires us to retrain or fine-tune some of the underlying NLP models in Marvista. Further, we will recruit more diverse users, e.g., those with low tech-savviness and various reading habits in order to validate whether Marvista could still provide value-added support for these readers.

---

[11]https://longreads.com/

### 7.2 Further Studies and Designs to Address Over-reliance on Marvista

It is worth noting that Marvista's current designs have already considered measures to discourage over-reliance, *e.g.*, showing an AI-generated summary only after a user indicates finishing reading the article. When designing the output of the time and question filters, instead of letting a reader skip to AI's recommend-to-read text, we chose to just highlight those text so that the reader still have equal access to all the text in the article. We believe this design choice reduces biasing readers with AI's recommendations. We also believe that explanations of the summary can contribute to reducing over-reliance because it allows the reader to question the AI's summary, *e.g.*, by checking whether a sentence in the summary might have come from unimportant parts of the article.

Connecting to prior work, Buçinca *et al.* summarized three design strategies to mitigate AI over-reliance [8]: *(i)* asking the person to make a decision before seeing the AI's recommendation, *(ii)* slowing down the process, and *(iii)* letting the person choose whether and when to see the AI recommendation. In the future, we plan to adopt these strategies in redesigning Marvista. For example, instead of showing the AI-generated summary right away, we can prompt the reader to first write their own summary or answer a few global reflective questions.

### 7.3 From Supporting Reading to Encouraging Reading

According to the American Time Use Survey [16], there is a slow and continuous decline in how much time Americans spend on reading on a daily basis. In 2019, individuals between 15 and 44 read on average for 10 minutes or less per day. By involving AI to collaborate with and support human readers, Marvista intends to lower the barrier of reading (long) text and eventually to encourage people to read more. In our future work, we plan to deploy Marvista in a longitudinal study, to track how much each participant reads, and to investigate whether using Marvista leads to more reading. After that, we also would like to study whether people would still read more without using Marvista—whether Marvista can boost one's intrinsic reading motivation or merely serve as an extrinsic assistive tool for reading more.

Building on Marvista as a representative example, future work can also explore the design space of human-AI collaborative reading tool to identify and lay out key design dimensions. Given how large language models are enabling a plethora of information processing tools, constructing a design space can systematically guide the development of such human-AI collaborative tools.

### 7.4 Implications for General Human-AI Collaboration Tools

We consider the main generalizable lesson from designing Marvista is that AI can collaborate with human knowledge workers by preprocessing information and using the generated insights (e.g., a summary or questions) to guide humans to optimize how they navigate the information (e.g., reading specific texts with targeted questions in mind). One analogous scenario to Marvista would be guiding physicians to navigate medical imaging data, e.g., prompting the physician to focus on regions of interest selected by an AI model or generating a summary of diagnosis that also considers the physician's attention and annotation on the data [23].

### 7.5 Technical Limitations of Marvista's Current Implementation

*7.5.1 A better measurement than dwelling time of each paragraph.* Currently we use dwelling time on each paragraph to gauge a reader's interest of that paragraph, which inevitably includes non-reading time. A more accurate measure would involve using eye tracking mechanisms to calculate the amount of time a reader is actually looking at the text.

*7.5.2 Exploring other types of summary explanations.* Currently, we employ a type of post hoc explanation [57] that compare the AI's summary with the sources (article text and user-generated data). One limitation of this approach is that it cannot reveal the 'black box' process of how a summary is actually generated. To achieve a more comprehensive explanation, our future work will explore and incorporate other types of explanations, *e.g.*, computing a saliency score of each paragraph (or even sentence) to reflect the actual process of how they are weightedly combined to formulate the summary.

*7.5.3 Supporting other reading platforms.* As shown in HCI literature (*e.g.*, [46]), people read on a variety of platforms: phones, tablets, laptop screens, and desktop monitors. Smart phones, in particular, have become the dominant platform for accessing all kinds of information including reading articles [40]. We are interested in redesigning Marvista as a smart phone app. Specifically, we envision a large vocabulary of input gestures and modalities are possible in the focus mode beyond conventionally scrolling the text up and down. Further, the presentation of the text can go beyond the traditional format, *e.g.*, as conversations between the reader and the AI where the two can ask and answer questions related to the text to enhance the interactivity of reading.

*7.5.4 Overcoming current limitations of NLP models.* Currently, our NLP models generate two types of summarizations. **Extractive summarization**: Directly selecting a few text subset from a given document can be simplified as a sentence ranking problem. Before a user reads an article, Marvista highlights several sentences based on the time or question filter. However, as mentioned by several participants, at times it can be hard to understand an extracted sentence without reading its context and if a user needs to read more context to understand our recommended sentences, then the time saved while reading is limited. In the future, Marvista can involve coreference resolution model to detect and annotate ambiguous entities. For example, it will be more understandable as an extractive sentence if we replace the "they" in the paragraph "they found that patients infected with the Omicron..." to "The South African Medical Research Council found ...".

**Abstractive summarization**: The abstractive summarization model used in Marvista is trained on the News articles because it has the richest human summary annotations. However, Marvista is a general tool that might be applied to any domain of text. There are two potential ways to mitigate this issue. The first is joint training with multi-domain summaries but it could be very costly to scale up. The other way is to conduct unsupervised reference-free pretraining such as masked-language modeling on the type of articles on which a reader focuses. Another major limitation of neural summarization models is the length of input and output that Marvista can digest. Existing generative language models has a fixed maximal number of input tokens such as 512 or 2048 tokens. Thus with extremely long articles Marvista can only take truncated article as input. In addition, the length of output summary and the granularity of the summary are highly depending on the training data annotation. For example, the summary annotated in XSUM [49] is shorter and more abstractive than the one annotated for the common CNN/Daily Mail corpus [48]. Lastly, the way Marvista incorporating user explicit or implicit feedback into summary generation could be one of the future study. In the current design, Marvista adjusts the encoder attention weights in a heuristic way without fine-tuning any model parameters.

Model hallucinations in abstractive summarization is another limitation. Currently, we mitigate such problems by adding an explainability feature to the abstractive summaries, thus enabling easier review by the user in cases where the abstractive summarizer might hallucinate. Hallucinations and factual inconsistencies in generated text remain an open research problem and we believe that we have put some guardrails in place to minimize risks, while allowing us to experiment with the new technology. In our future work, we can add hallucination detection models to filter or rerank summaries (*e.g.*, the Qafacteval approach [19]), or use post-processing methods (*e.g.*, [18]), or even improve model training (*e.g.*, [15]).

**Question generation**: The existing generated questions might focus too much on low-level facts and do not capture the high-level ideas of the text. The primary reason is that the training QA corpora are mostly designed for the machine reading comprehension task, which usually contains short and factual answers. To generate more meaningful questions beyond *remembering*, our future work involves training models that produce questions at the level of *understanding*, *applying*, *analyzing*, *evaluating*, and *creating*.

## 8 CONCLUSION

As humans, our ability to read has not caught up with the growing amount of information available for us to consume. Just as Herbert A. Simon once said, "... a wealth of information creates a poverty of attention". Marvista represents a mission of creating human-AI collaborative tools to help us overcome "a poverty of attention" in the ever-increasing "information-rich world". Our study participants not only benefited from using Marvista to help them read more efficiently, focusedly, and informedly, via interacting with Marvista they also learned more about their own reading behavior. As we continue to imporve the design and AI models of Marvista, we hope to employ this tool to encourage more people to read more and to enhance both our ability and interest in thriving in this information-rich world.

Marvista: Exploring the Design of a Human-AI Collaborative News Reading Tool 27*Technology* (Virtual Event, USA) *(UIST '20 Adjunct)*. Association for Computing Machinery, New York, NY, USA, 132–134. https://doi.org/10.1145/3379350.3416154
[56] Horst Pöttker. 2003. News and its communicative quality: the inverted pyramid—when and why did it appear? *Journalism Studies* 4, 4 (2003), 501–511.
[57] Forough Poursabzi-Sangdeh, Daniel G Goldstein, Jake M Hofman, Jennifer Wortman Wortman Vaughan, and Hanna Wallach. 2021. Manipulating and measuring model interpretability. In *Proceedings of the 2021 CHI Conference on Human Factors in Computing Systems*. 1–52.
[58] Morgan N Price, Bill N Schilit, and Gene Golovchinsky. 1998. XLibris: The active reading machine. In *CHI 98 conference summary on Human factors in computing systems*. 22–23.
[59] Colin Raffel, Noam Shazeer, Adam Roberts, Katherine Lee, Sharan Narang, Michael Matena, Yanqi Zhou, Wei Li, and Peter J Liu. 2019. Exploring the limits of transfer learning with a unified text-to-text transformer. *arXiv preprint arXiv:1910.10683* (2019).
[60] Francis P Robinson. 1946. Effective study, Rev. (1946).
[61] D Shaikh. 2004. Paper or pixels: What are people reading online. *Usability News* 6, 2 (2004).
[62] Alexandra N Spichtig, Elfrieda H Hiebert, Christian Vorstius, Jeffrey P Pascoe, P David Pearson, and Ralph Radach. 2016. The decline of comprehension-based silent reading efficiency in the United States: A comparison of current data with performance in 1960. *Reading Research Quarterly* 51, 2 (2016), 239–259.
[63] Craig S Tashman and W Keith Edwards. 2011. LiquidText: A flexible, multitouch environment to support active reading. In *Proceedings of the SIGCHI Conference on Human Factors in Computing Systems*. 3285–3294.
[64] Loren G. Terveen. 1995. Overview of human-computer collaboration. *Knowledge-Based Systems* 8, 2 (1995), 67–81. https://doi.org/10.1016/0950-7051(95)98369-H Human-computer collaboration.
[65] Mark Thomas. 2020. What is the average reading speed and the best rate of reading? https://www.healthguidance.org/entry/13263/1/What-Is-the-Average-Reading-Speed-and-the-Best-Rate-of-Reading.html
[66] Steve Tsang, Ravin Balakrishnan, Karan Singh, and Abhishek Ranjan. 2004. A Suggestive Interface for Image Guided 3D Sketching. In *Proceedings of the SIGCHI Conference on Human Factors in Computing Systems* (Vienna, Austria) *(CHI '04)*. Association for Computing Machinery, New York, NY, USA, 591–598. https://doi.org/10.1145/985692.985767
[67] Philipp Tschandl, Christoph Rinner, Zoe Apalla, Giuseppe Argenziano, Noel Codella, Allan Halpern, Monika Janda, Aimilios Lallas, Caterina Longo, Josep Malvehy, John Paoli, Susana Puig, Cliff Rosendahl, H. Peter Soyer, Iris Zalaudek, and Harald Kittler. 2020. Human–computer collaboration for skin cancer recognition. *Nature Medicine* 26, 8 (Aug. 2020), 1229–1234. https://doi.org/10.1038/s41591-020-0942-0 Number: 8 Publisher: Nature Publishing Group.
[68] Nobuyuki Umetani, Takeo Igarashi, and Niloy J. Mitra. 2012. Guided Exploration of Physically Valid Shapes for Furniture Design. *ACM Trans. Graph.* 31, 4, Article 86 (jul 2012), 11 pages. https://doi.org/10.1145/2185520.2185582
[69] Ashish Vaswani, Noam Shazeer, Niki Parmar, Jakob Uszkoreit, Llion Jones, Aidan N Gomez, Łukasz Kaiser, and Illia Polosukhin. 2017. Attention is all you need. In *Advances in neural information processing systems*. 5998–6008.
[70] Dakuo Wang, Justin D. Weisz, Michael Muller, Parikshit Ram, Werner Geyer, Casey Dugan, Yla Tausczik, Horst Samulowitz, and Alexander Gray. 2019. Human-AI Collaboration in Data Science: Exploring Data Scientists' Perceptions of Automated AI. *Proc. ACM Hum.-Comput. Interact.* 3, CSCW, Article 211 (Nov. 2019), 24 pages. https://doi.org/10.1145/3359313
[71] Nora S. Willett, Rubaiat Habib Kazi, Michael Chen, George Fitzmaurice, Adam Finkelstein, and Tovi Grossman. 2018. A Mixed-Initiative Interface for Animating Static Pictures. In *Proceedings of the 31st Annual ACM Symposium on User Interface Software and Technology* (Berlin, Germany) *(UIST '18)*. Association for Computing Machinery, New York, NY, USA, 649–661. https://doi.org/10.1145/3242587.3242612
[72] Songhua Xu, Hao Jiang, and Francis C.M. Lau. 2009. User-Oriented Document Summarization through Vision-Based Eye-Tracking. In *Proceedings of the 14th International Conference on Intelligent User Interfaces* (Sanibel Island, Florida, USA) *(IUI '09)*. Association for Computing Machinery, New York, NY, USA, 7–16. https://doi.org/10.1145/1502650.1502656
[73] Dongwook Yoon, Ken Hinckley, Hrvoje Benko, François Guimbretière, Pourang Irani, Michel Pahud, and Marcel Gavrilu. 2015. Sensing tablet grasp+ micro-mobility for active reading. In *Proceedings of the 28th Annual ACM Symposium on User Interface Software & Technology*. 477–487., Vol. 1, No. 1, Article . Publication date: June 2018.